\documentclass{article}

\usepackage{PRIMEarxiv}

\usepackage[utf8]{inputenc} % allow utf-8 input
\usepackage[T1]{fontenc}    % use 8-bit T1 fonts
\usepackage{hyperref}       % hyperlinks
\usepackage{url}            % simple URL typesetting
\usepackage{booktabs}       % professional-quality tables
\usepackage{amsfonts}       % blackboard math symbols
\usepackage{nicefrac}       % compact symbols for 1/2, etc.
\usepackage{microtype}      % microtypography
\usepackage{lipsum}
\usepackage{fancyhdr}       % header
\usepackage{graphicx}       % graphics
\usepackage{subfig}
\graphicspath{{media/}}     % organize your images and other figures under media/ folder

\title{Sampling color and geometry point clouds from ShapeNet dataset
}

\author{
  Davi Lazzarotto and Touradj Ebrahimi \\
  Multimedia Signal Processing Group (MMSPG) \\
  École Polytechnique Fédérale de Lausanne (EPFL) \\
  Lausanne, Switzerland\\
  \texttt{davi.nachtigalllazzarotto@epfl.ch, touradj.ebrahimi@epfl.ch} \\
}

\begin{document}
\maketitle

\begin{abstract}

The popularisation of acquisition devices capable of capturing volumetric information such as LiDAR scans and depth cameras has lead to an increased interest in point clouds as an imaging modality. 
Due to the high amount of data needed for their representation, efficient compression solutions are needed to enable practical applications. 
Among the many techniques that have been proposed in the last years, learning-based methods are receiving large attention due to their high performance and potential for improvement. 
Such algorithms depend on large and diverse training sets to achieve good compression performance. 
ShapeNet is a large-scale dataset composed of CAD models with texture and constitute and effective option for training such compression methods. 
This dataset is entirely composed of meshes, which must go through a sampling process in order to obtain point clouds with geometry and texture information. 
Although many existing software libraries are able to sample geometry from meshes through simple functions, obtaining an output point cloud with geometry and color of the external faces of the mesh models is not a straightforward process for the ShapeNet dataset.
The main difficulty associated with this dataset is that its models are often defined with duplicated faces sharing the same vertices, but with different color values. 
This document describes a script for sampling the meshes from ShapeNet that circumvent this issue by excluding the internal faces of the mesh models prior to the sampling. 
The script can be accessed from the following link: \url{https://github.com/mmspg/mesh-sampling}.

\end{abstract}

% keywords can be removed

\section{Scope and Background}

The development of imaging modalities for the representation of three-dimensional content has been an important topic of research in the last decades. 
The increasing performance of computing devices together with the high quality of modern displays have allowed for a fast development of the field of computer graphics for both industrial and entertainment applications to mention two among a large number of potential applications. 
Traditionally, this field has relied on the use of meshes as the imaging modality for the representation of artificially generated content. 

Mesh models are represented as a set of interconnected points in the three dimensional space. 
These vertices and edges define a set polygons that usually constitute the surface of a watertight volume. 
On one hand, the color on the faces of such 3D models can be defined either as values assigned individually for each face or as two dimensional texture, mapped directly onto the surface. 
Point clouds, on the other hand, don't contain any connectivity information, being composed uniquely, of a list of point coordinates with associated attributes such as color, normal vectors, semantic labels and many others possible features. 

The advent and popularization of acquisition devices that allow to capture volumetric information such as LiDAR scans and depth cameras has fostered the rise of new applications such as telepresence, virtual reality and wide area scanning. 
The output of such devices can usually be easily converted into a list of the space coordinates of the acquired points with associated attributes such as color and reflectance. 
Although there are algorithms capable of generating meshes from the scans, in many applications it is more advantageous to directly use of the acquired points in the form of point clouds. 

Depending on application, the number of points in a typical point cloud model can range from thousands up to the order of billions. 
Since the transmission and storage of such huge amount of data is impractical, efficient compression methods are paramount. 
For this reason, standardisation committees such as JPEG, Khronos Group and MPEG have been devoting efforts to the development of interoperable compression standards.

\section{Current practices and challenges}

While many conventional data structures such as the octree or the sets of projections have been proposed to encode point cloud data, deep learning-based architectures have been reporting high performance and have attracted the attention of many researchers and standardisation groups. 
Such methods apply transforms learned through a training process, relying on large and diverse datasets with thousands of point clouds. 
Several datasets were employed to train learning-based compression algorithms reported in the literature. 
Table~\ref{tab:codecs_datasets} lists some of these datasets including the reference to the respective compression methods.

\begin{table}[]
    \centering
    \begin{tabular}{c|c}
        Dataset & Compression method\\ \hline
        ShapeNet\cite{shapenet2015}  & Wang et al.\cite{Wang2020b, Wang2021a}\\
        ModelNet\cite{modelnet2015}  & Quach et al.\cite{Quach2019a,Quach2020a} and Nguyen et al.\cite{Nguyen2021a} \\
        MPEG & Alexiou et al.\cite{Alexiou2020a} and Guarda et al.\cite{Guarda2019a,Guarda2019b,Guarda2020a,Guarda2020b,Guarda2020c,Guarda2020d,Guarda2020e} \\
        JPEG Pleno\cite{jpeg_pleno} & Alexiou et al.\cite{Alexiou2020a} \\
        nuScenes\cite{nuscenes} & Wiesman et al.\cite{Wiesmann2021}
    \end{tabular}
    \caption{Datasets employed for training learning-based point cloud compression methods.}
    \label{tab:codecs_datasets}
\end{table}

Among the datasets listed in Table~\ref{tab:codecs_datasets}, ShapeNet is a powerful option for training learning-based compression methods due to its large number of models with associated color texture. 
Moreover, it has been already successfully employed for training geometry-only compression algorithms. 
Since ShapeNet is composed of mesh models, a preprocessing step is needed where in order to convert the dataset into point clouds prior to its use in the training loop. 
Although ignoring the connectivity information and forming a point cloud with the mesh vertices is in theory a possible solution, the resulting models will potentially have too low point density. 
Previous works~\cite{Quach2019a,Quach2020a, Wang2021a} used random sampling followed by voxelization in order to obtain geometry-only point clouds with points lying on a uniform grid. 
However, the software libraries by these authors are only capable of sampling the geometry, ignoring associated color attributes.

\section{Mesh sampling solutions}

\subsection{Software libraries}

Several~\cite{Quach2019a,Quach2020a,Wang2020b,Wang2021a,Alexiou2020a} open source learning-based point cloud compression methods are based on Python~\cite{python} as a programming language. 
Similarly, many software libraries for point cloud processing are based on Python as well, such as Pyntcloud~\cite{pyntcloud}, Open3D~\cite{Open3D} and pymeshlab~\cite{MeshLab}. 
Pyntcloud~\cite{pyntcloud} allows for the sampling of meshes through the method \verb-get_sample()-, which randomly selects a defined number of points from a mesh. 
This library was employed by the authors of~\cite{Quach2019a,Quach2020a, Wang2021a}, but is only able to generate geometry-only point clouds. 

The Open3D~\cite{Open3D} Python library has two methods for mesh sampling: \verb-sample_points_uniformly()- applies uniform sampling, while \verb-sample_points_poisson_disk()- uses Poisson disk sampling~\cite{poisson_disk_sampling} to obtain a point cloud from the mesh. Likewise, these methods are only able to deal with geometry-only data. 

Meshlab~\cite{MeshLab} is a standalone software that allows for the visualisation and processing of meshes and point clouds. It contains a large number of different algorithms for mesh sampling, which are also unable to generate point clouds with texture. Meshlab also has functions that transfer the color from mesh vertices to a point cloud, which are however not able to deal with cases where the color is defined as a two dimensional texture map. All functions from Meshlab are also available in a corresponding Python library called pymeshlab. 

CloudCompare~\cite{cloudcompare} is another tool that can be used for visualisation and processing of 3D content. This software contains a function that allows for the direct sampling of both the color and the geometry in random positions over a mesh surface. Moreover, is is able to deal with color defined both per face or as texture.

\subsection{Proposed method}

While CloudCompare can be used to sample point clouds with color, the direct sampling of mesh models from ShapeNet using this tool doesn't generate point clouds with similar visual result as the corresponding mesh renderings. 
Figure \ref{fig:original_mesh_bad_sampling} portrays this result, where it is clear that the color on some parts of the generated point cloud is corrupted, with the appearance of added noise. 
The reason for this effect is that meshes from ShapeNet are often defined with two or more faces sharing the same vertices, but with opposing normal vectors. 
Moreover, these pairs of faces at the same position in space don't always have the same color. 
Most mesh rendering software apply back face culling, not showing any faces whose normal is not facing the direction of visualization. 
However, the obtained point cloud is randomly sampled from all faces, and therefore points with different colors appear on the surface of the same face. 

\begin{figure}
    \subfloat[Example mesh from ShapeNet dataset]{\begin{minipage}[b]{0.5\linewidth}
    \centering
    \includegraphics[width=\linewidth]{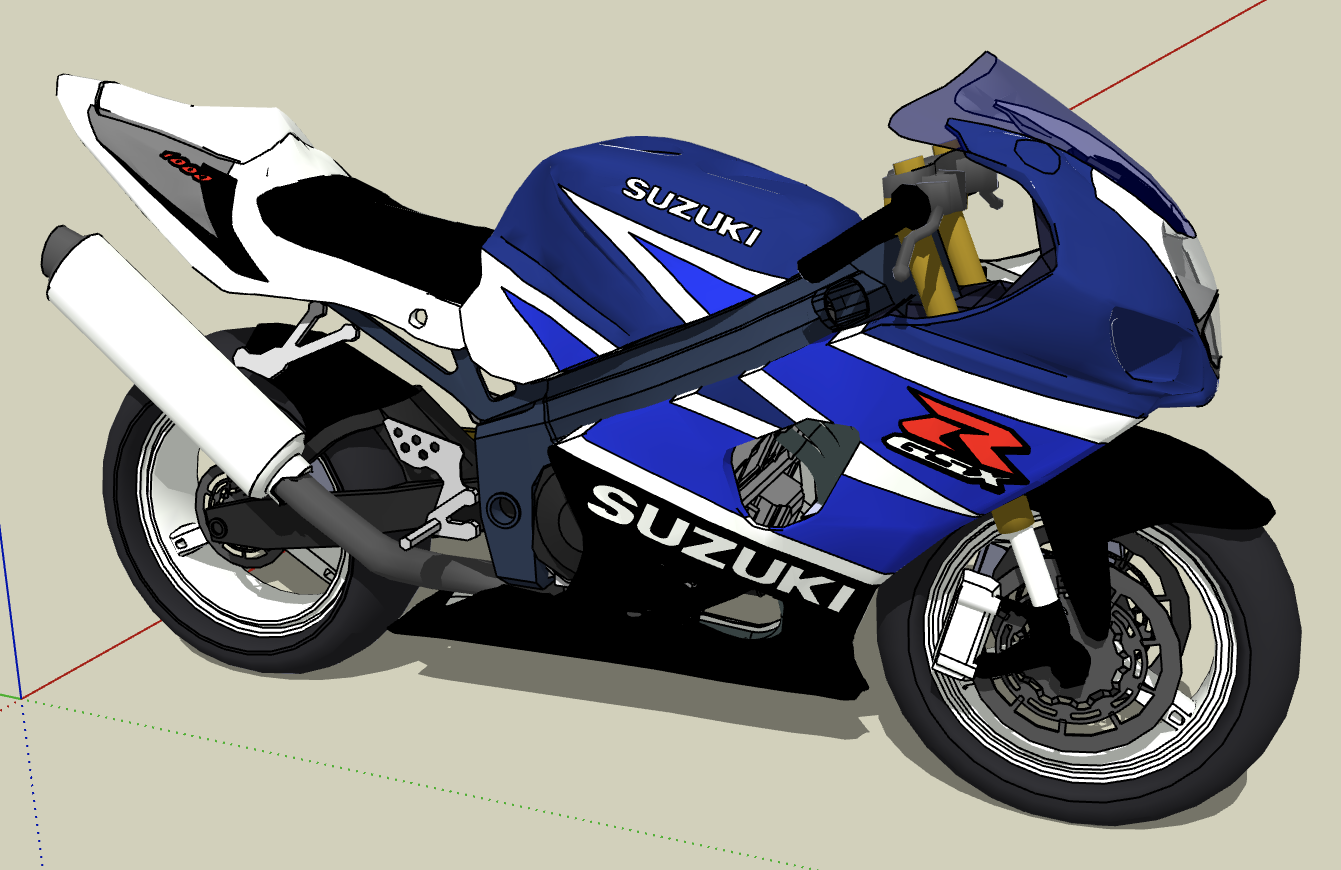}
    \label{fig:original_mesh}
    \end{minipage}}
    \subfloat[Point cloud generated after direct sampling]{
    \begin{minipage}[b]{0.5\linewidth}
    \centering
    \includegraphics[width=\linewidth]{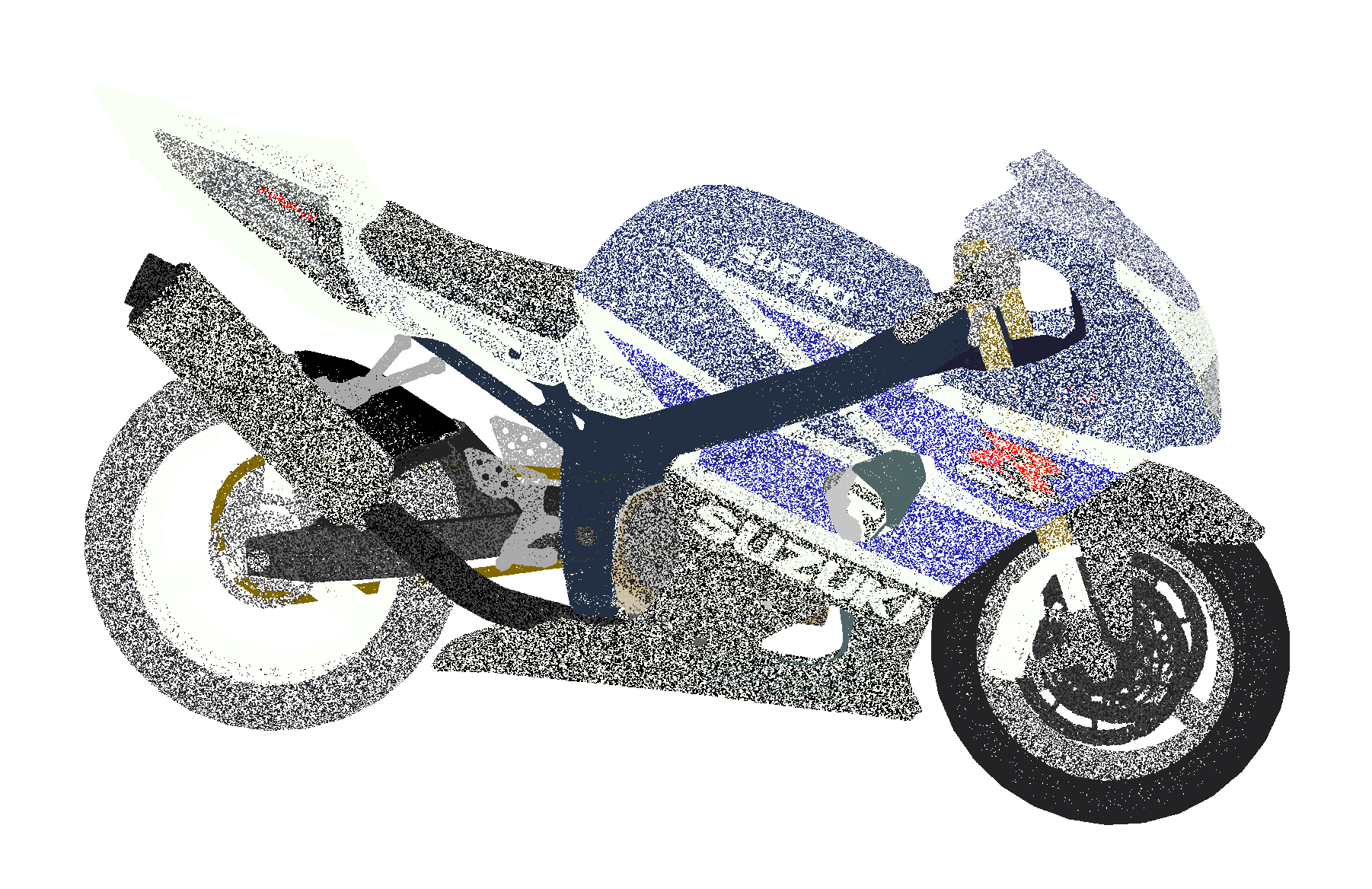}
    \label{fig:bad_sampling}
    \end{minipage}}
    \caption{Example mesh \emph{SuzukiGSXR10002003} and corresponding direct sampling}
    \label{fig:original_mesh_bad_sampling}
\end{figure}

In order to circumvent this issue, our proposed method aims to detect which faces have normal vectors facing the interior of the mesh and cannot be seen from the position of an outside observer. 
These faces can be then removed from the mesh prior to sampling so that their colors don't affect the generated point cloud. 
This process is done through the ambient occlusion plugin from Meshlab, which simulates multiple view directions around the mesh model and assigns to each face a quality value proportional to the number of views from each that face is visible. 
Then, an iterative process scans the mesh identifying the faces that share the same vertices and selecting only the faces with highest quality value among them and removing the remaining ones. 
After this process, only the faces with normal vectors facing the exterior of the mesh should be present, and no duplicated faces in the same position remain. 
The produced mesh is then sampled using CloudCompare, generating a point cloud with both color and geometry. 

Finally, the point cloud is scaled to a cubic bounding box with a desired resolution and the point coordinates are quantized to an uniform grid. If two or more points are quantized to the same coordinate, the associated color attributes are average out. This process is applied using the Open3D library. The result for the exemple mesh model can be seen in Figure~\ref{fig:script_pc}.

\begin{figure}

    \centering
    \includegraphics[width=0.5\textwidth]{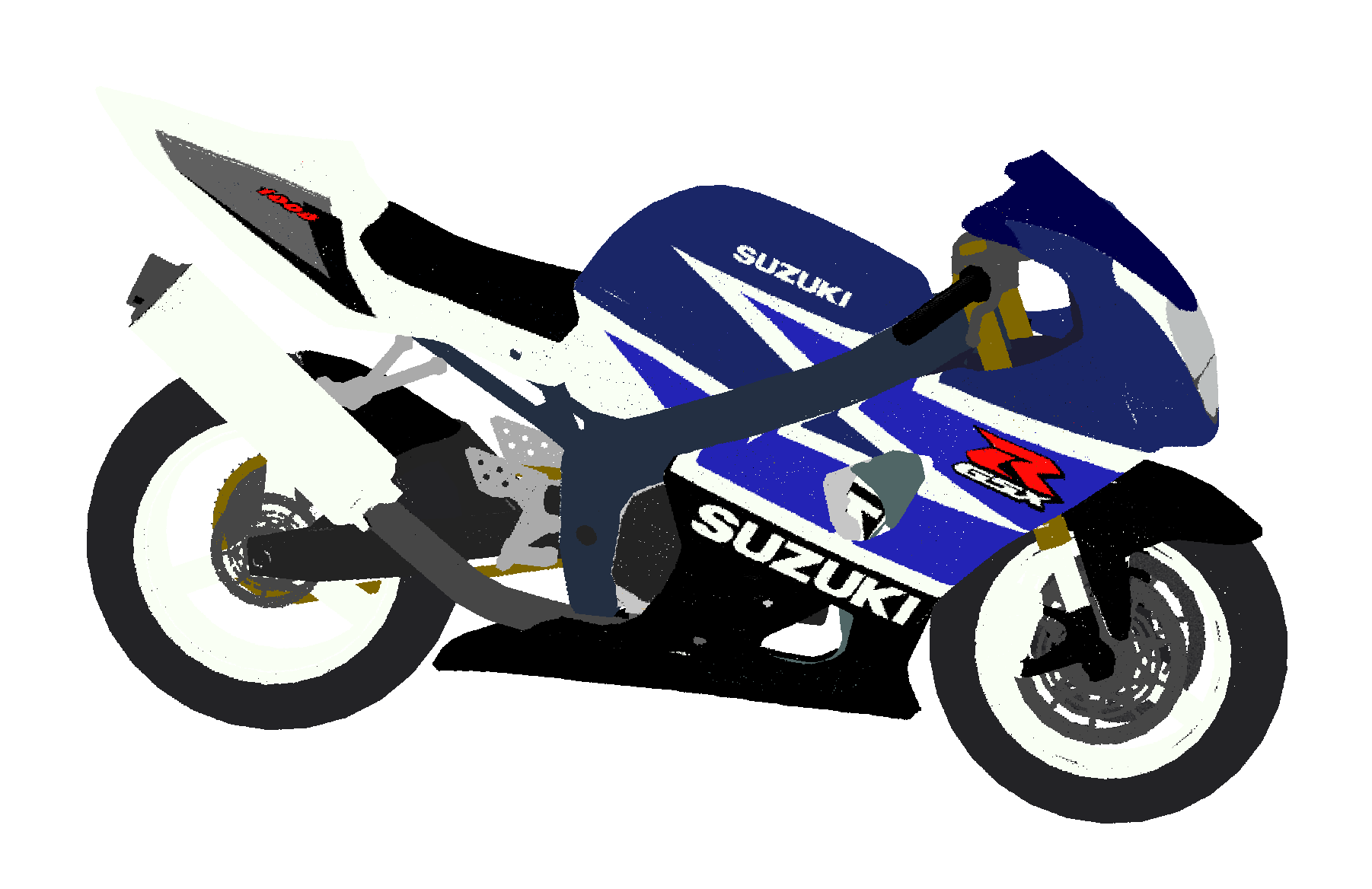}
    \caption{Point cloud generated from \emph{SuzukiGSXR10002003} by proposed script}
    \label{fig:script_pc}
\end{figure}

\section{Condition of use}

If you wish to use the provided script in your research, we kindly ask you to cite this document. 

\section*{Acknowledgments}
This work was supported by the Swiss National Foundation for Scientific Research (SNSF) under the grant number 200021-178854.

%Bibliography
\bibliographystyle{unsrt}  
\bibliography{references}

\appendix

\section{Annex A}

In this annex, there are examples of the application of the proposed scripts from other metrics from ShapeNet. 

\begin{figure}[h]
    \subfloat[Example mesh from ShapeNet dataset]{\begin{minipage}[b]{0.5\linewidth}
    \centering
    \includegraphics[width=\linewidth]{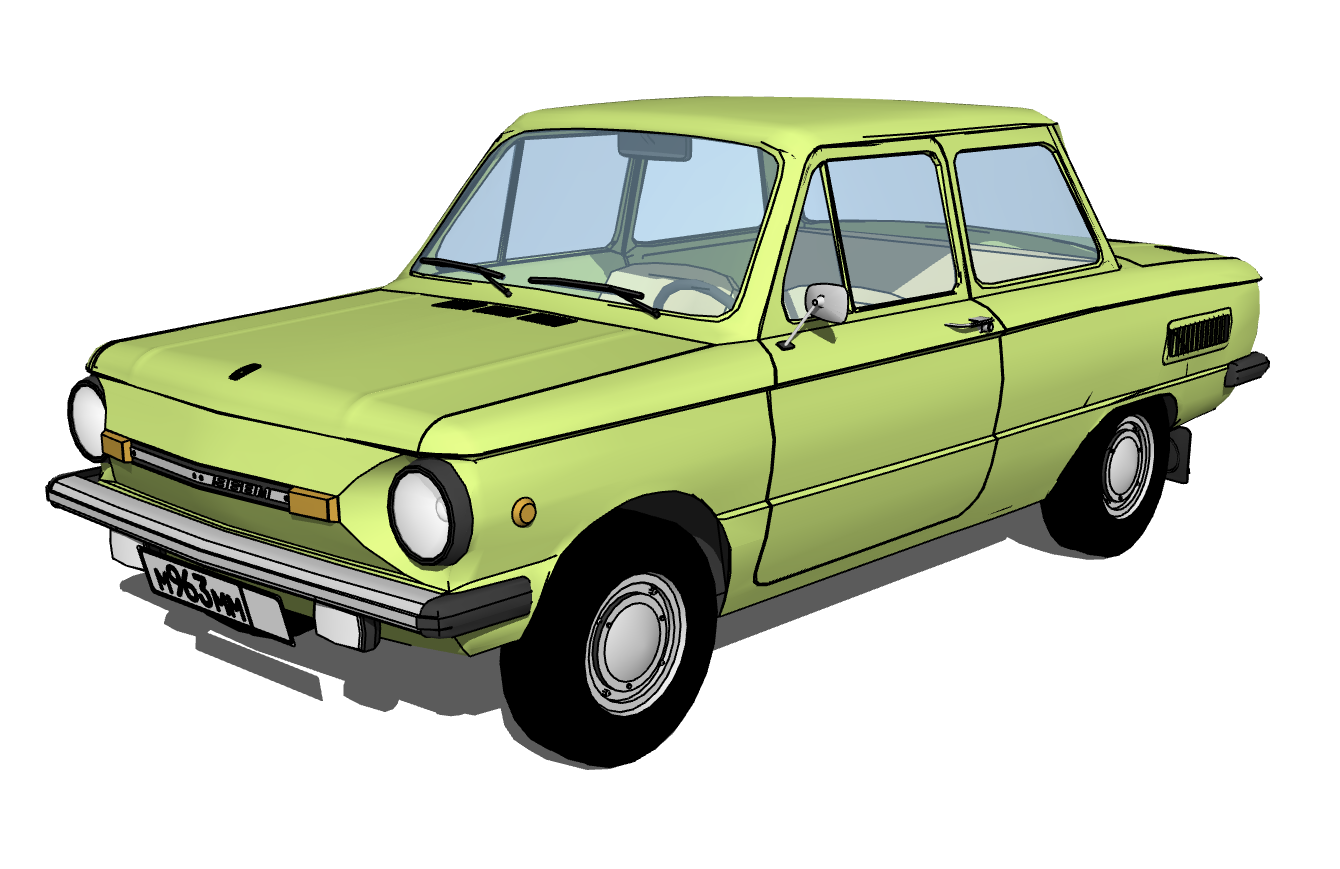}
    \label{fig:original_mesh}
    \end{minipage}}
    \subfloat[Point cloud generated after direct sampling]{
    \begin{minipage}[b]{0.5\linewidth}
    \centering
    \includegraphics[width=\linewidth]{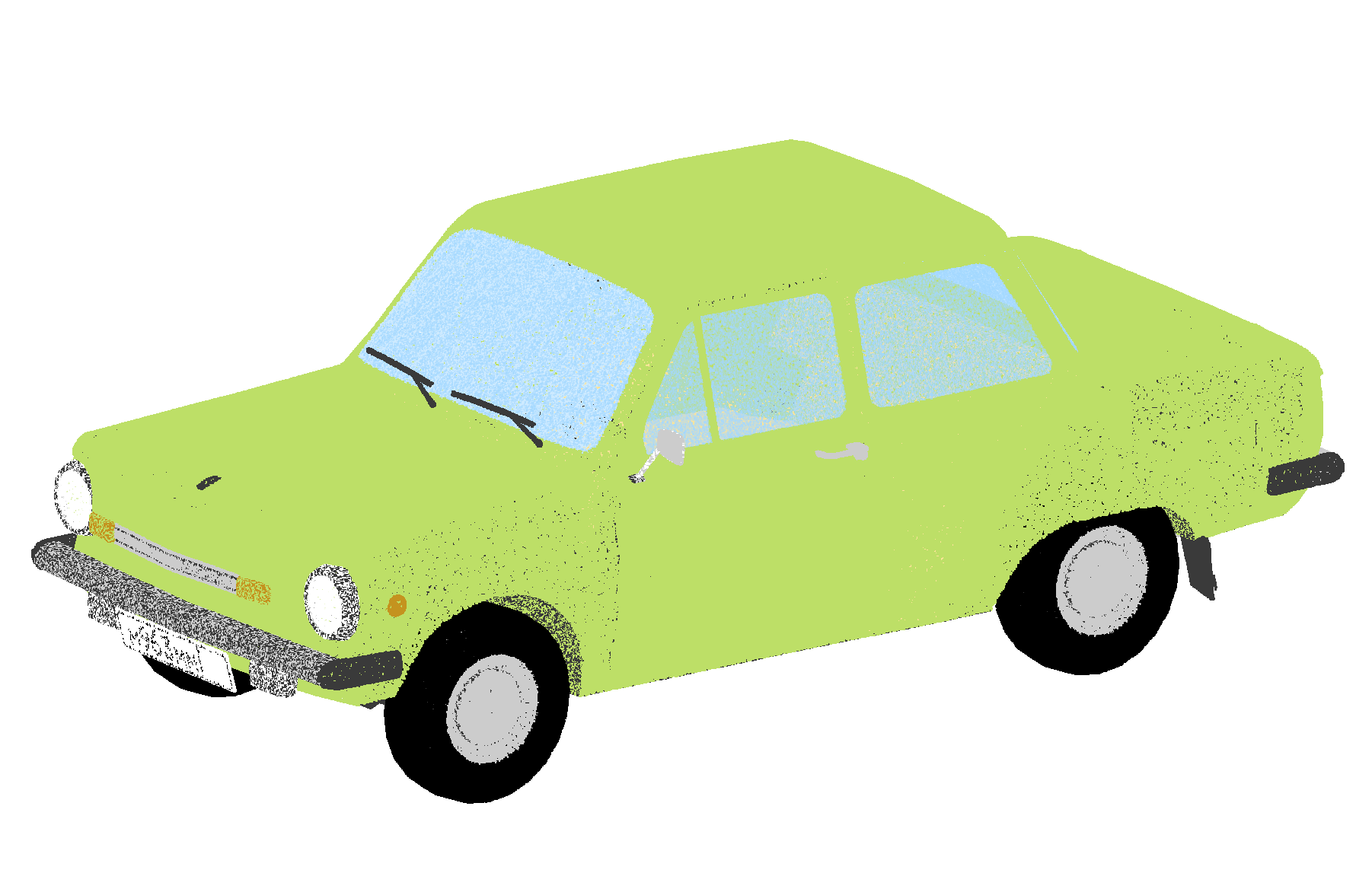}
    \label{fig:bad_sampling}
    \end{minipage}}
    \caption{Example mesh \emph{ZAZ968MZaporozhec} and corresponding direct sampling}
    \label{fig:original_mesh_bad_sampling}
\end{figure}

\begin{figure}[h]
    \centering
    \includegraphics[width=0.5\textwidth]{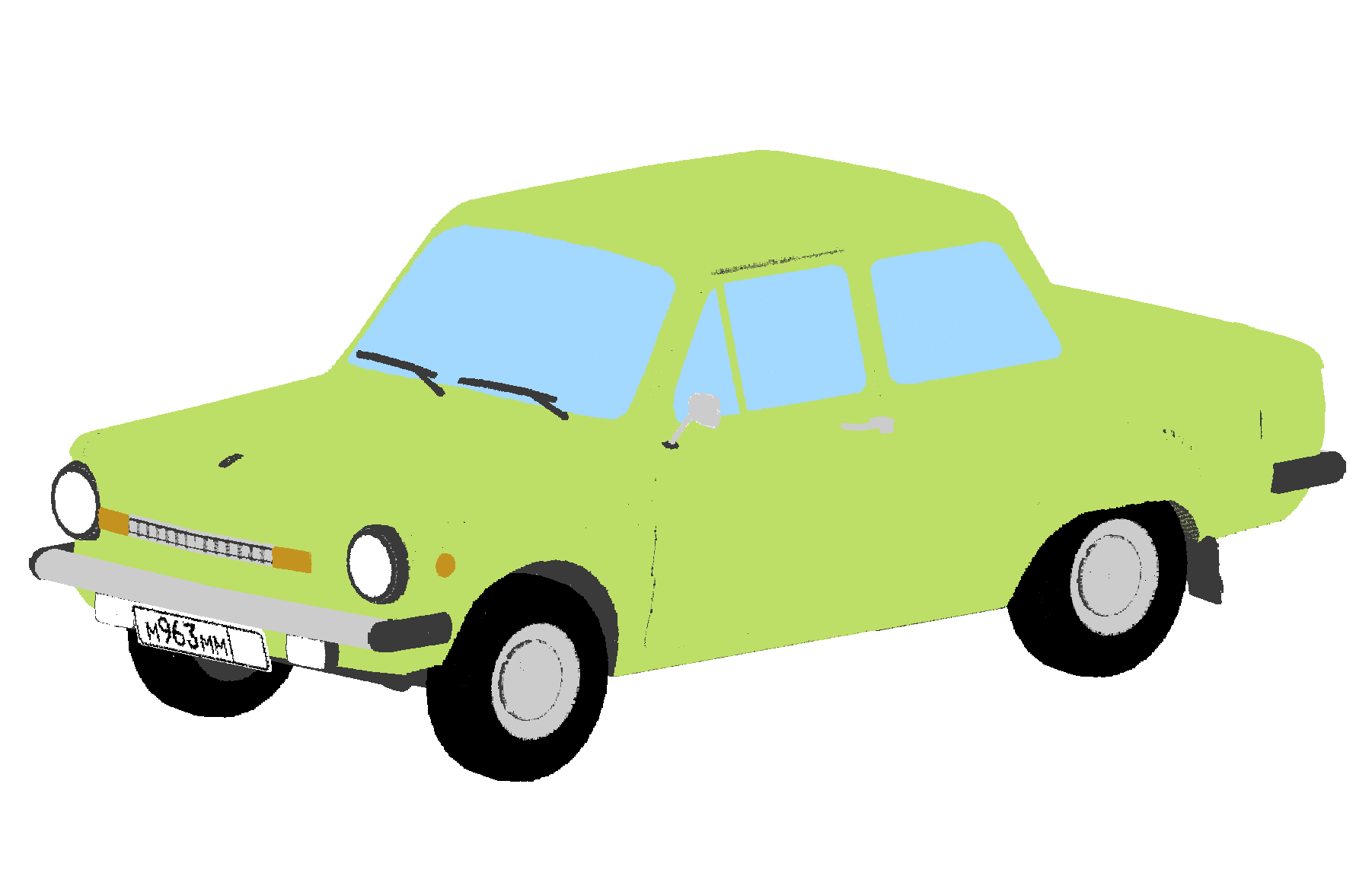}
    \caption{Point cloud generated from \emph{ZAZ968MZaporozhec} by proposed script}
    \label{fig:script_pc}
\end{figure}

\begin{figure}[h]
    \subfloat[Example mesh from ShapeNet dataset]{\begin{minipage}[b]{0.5\linewidth}
    \centering
    \includegraphics[width=\linewidth]{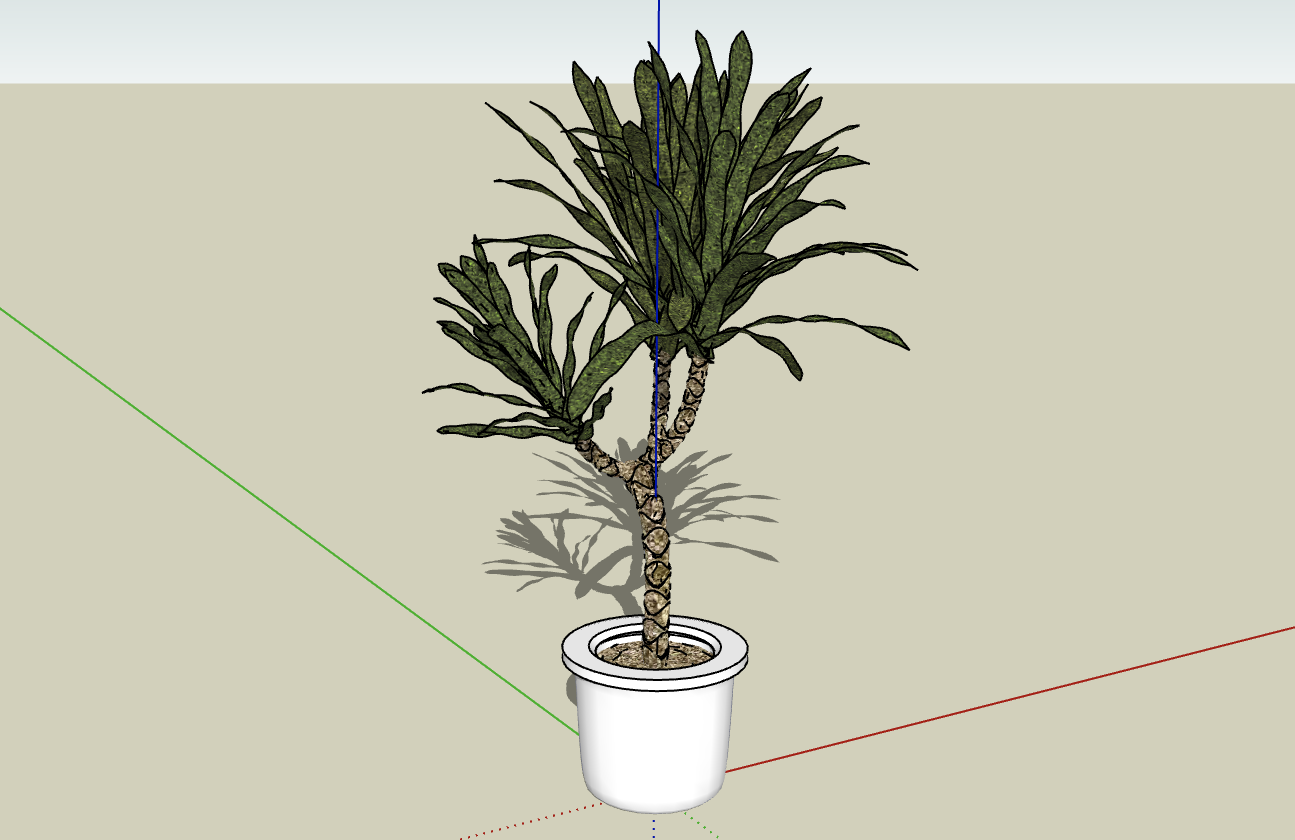}
    \label{fig:original_mesh}
    \end{minipage}}
    \subfloat[Point cloud generated after direct sampling]{
    \begin{minipage}[b]{0.5\linewidth}
    \centering
    \includegraphics[width=\linewidth]{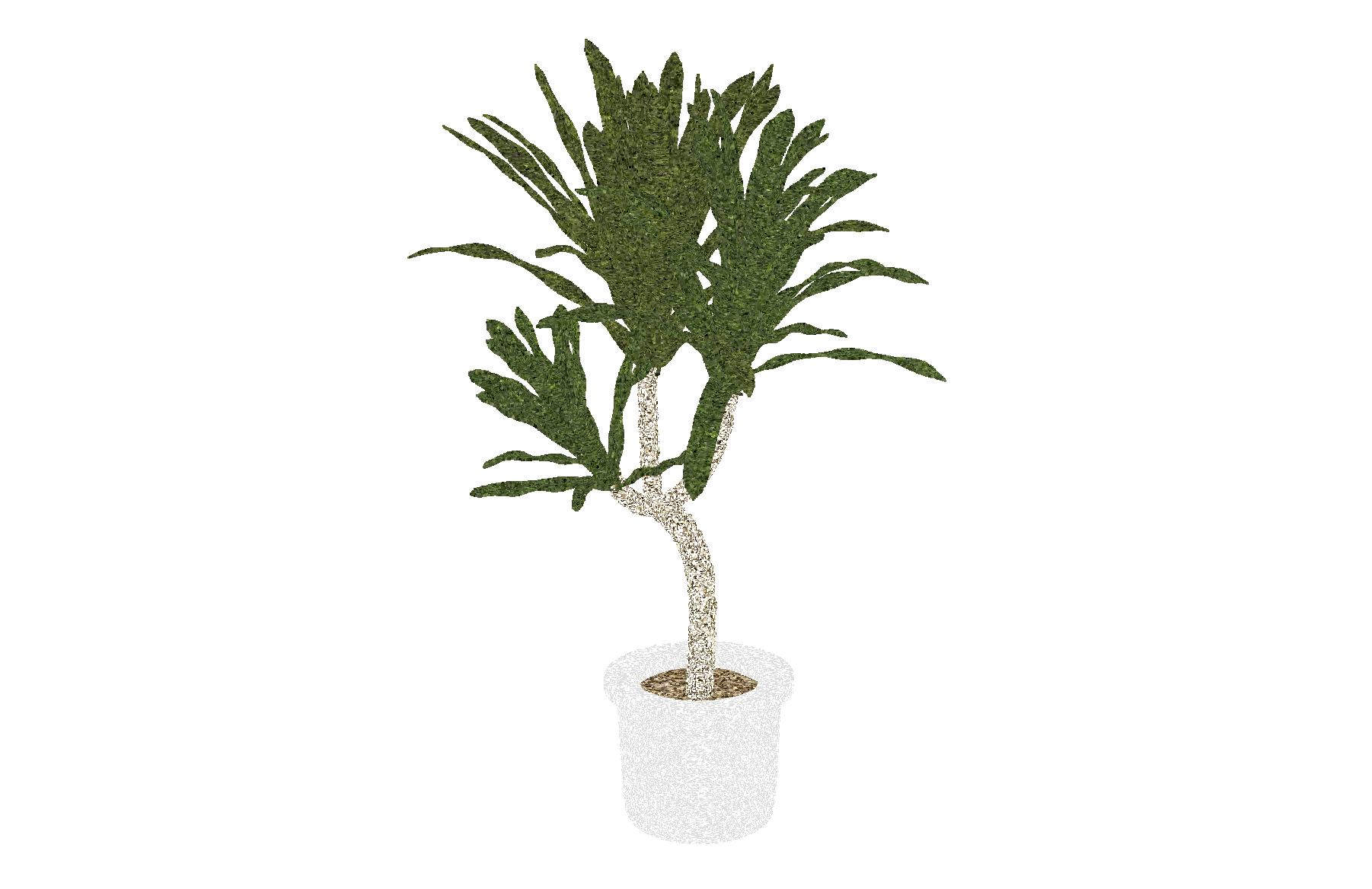}
    \label{fig:bad_sampling}
    \end{minipage}}
    \caption{Example mesh \emph{Plantanopote2} and corresponding direct sampling}
    \label{fig:original_mesh_bad_sampling}
\end{figure}

\begin{figure}[h]
    \centering
    \includegraphics[width=0.5\textwidth]{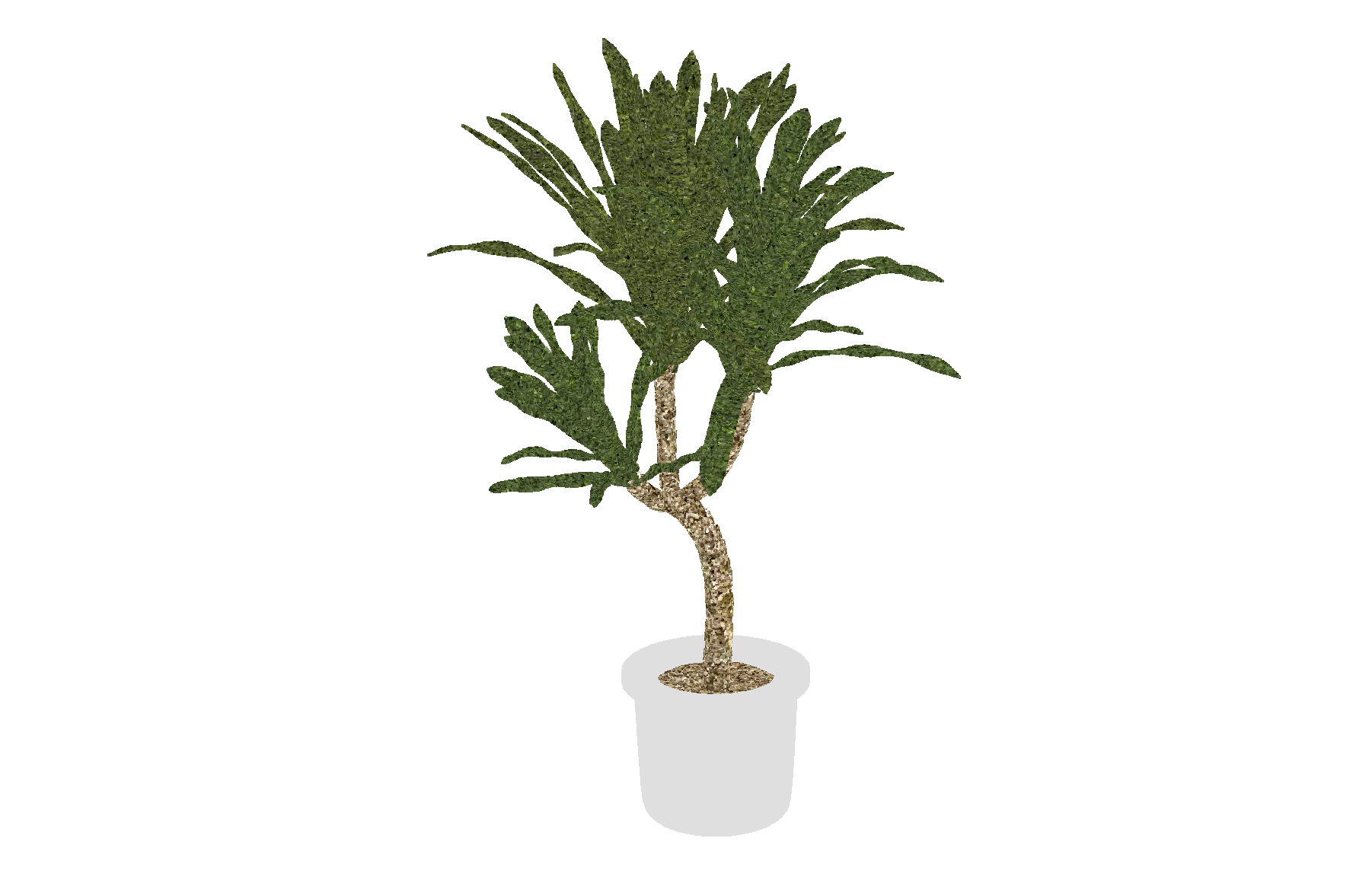}
    \caption{Point cloud generated from \emph{Plantanopote2} by proposed script}
    \label{fig:script_pc}
\end{figure}

\begin{figure}[h]
    \subfloat[Example mesh from ShapeNet dataset]{\begin{minipage}[b]{0.5\linewidth}
    \centering
    \includegraphics[width=\linewidth]{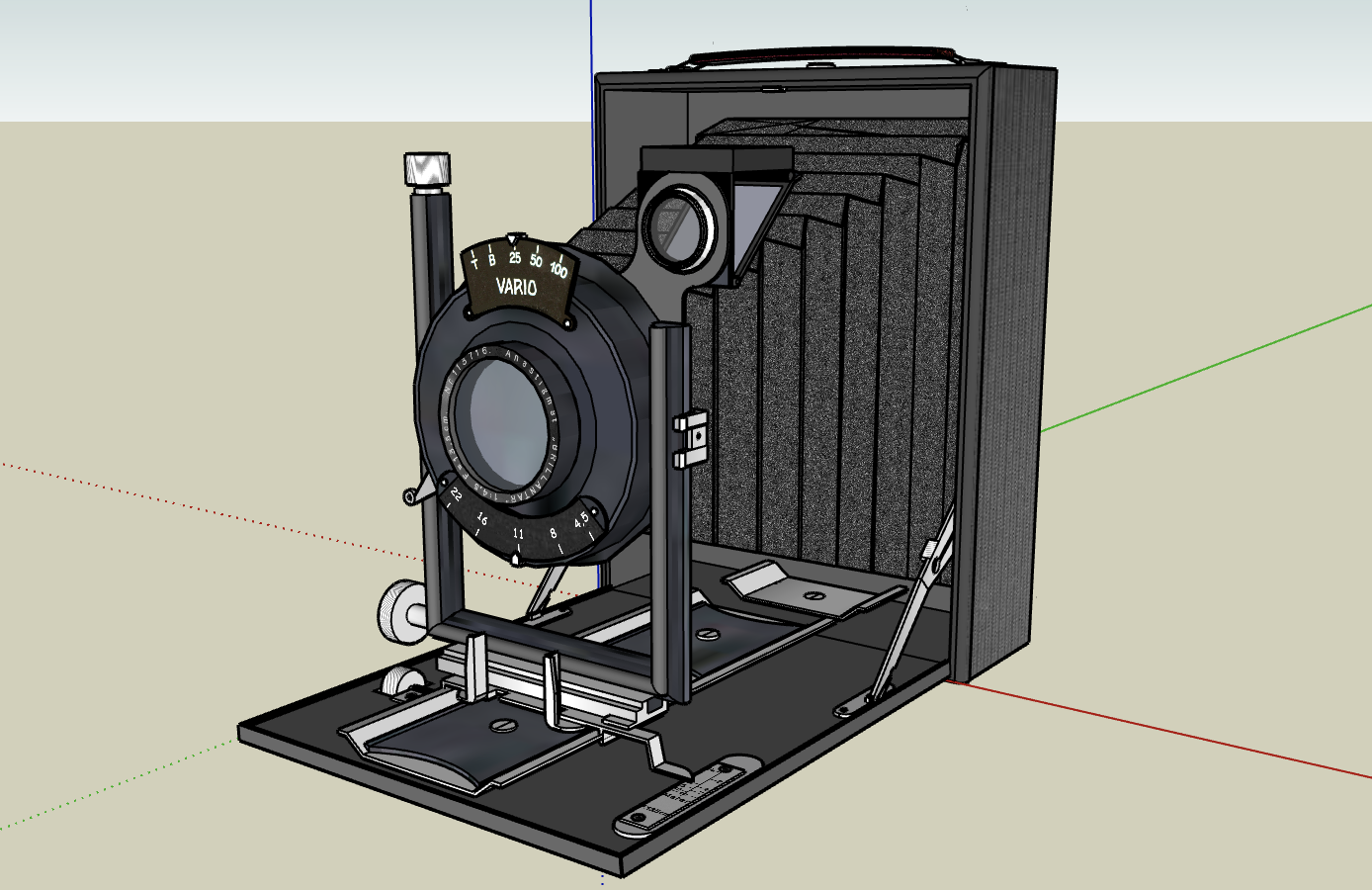}
    \label{fig:original_mesh}
    \end{minipage}}
    \subfloat[Point cloud generated after direct sampling]{
    \begin{minipage}[b]{0.5\linewidth}
    \centering
    \includegraphics[width=\linewidth]{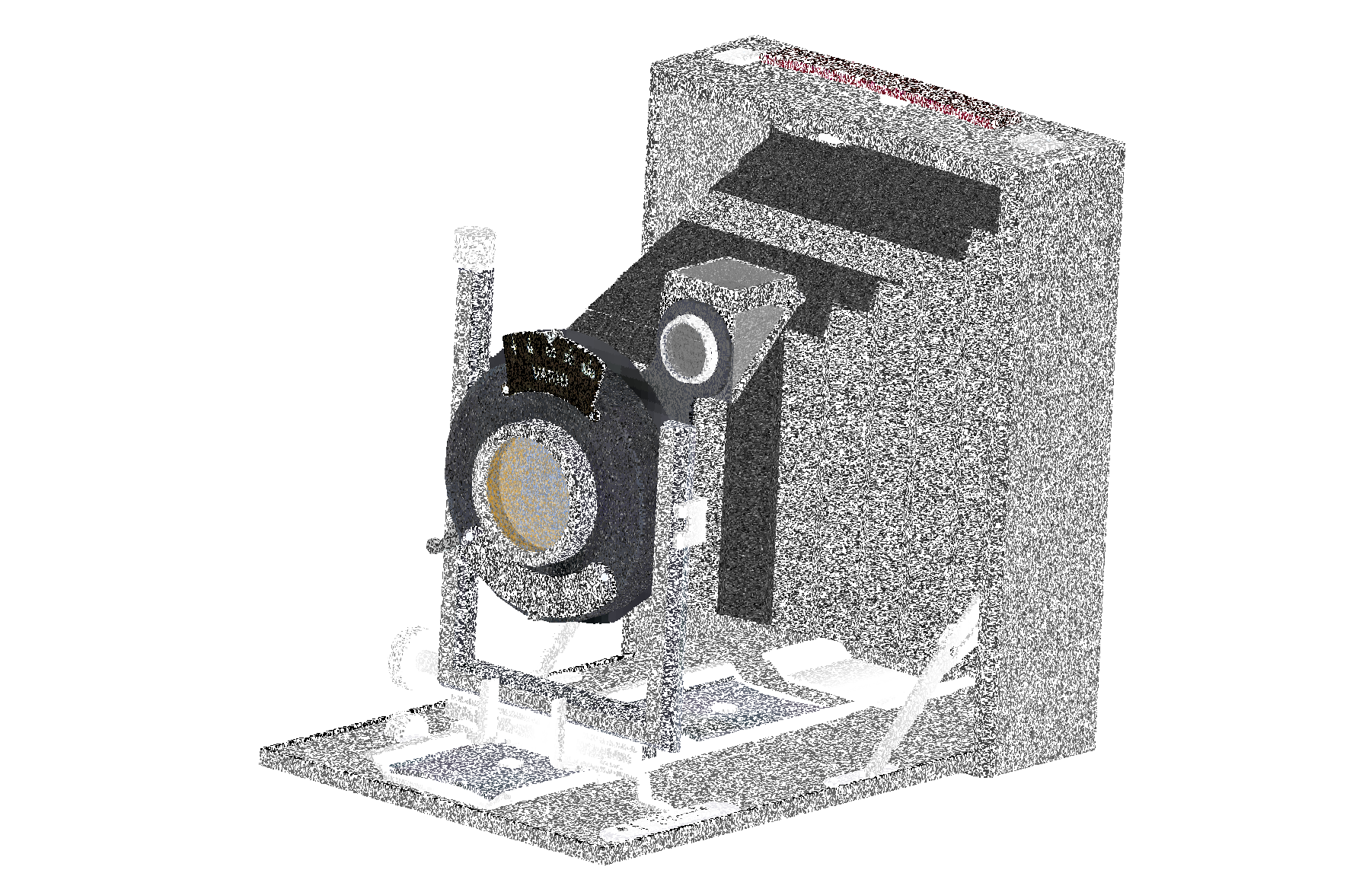}
    \label{fig:bad_sampling}
    \end{minipage}}
    \caption{Example mesh \emph{CameraVarioPlatterncamera} and corresponding direct sampling}
    \label{fig:original_mesh_bad_sampling}
\end{figure}

\makeatletter
\setlength{\@fptop}{0pt}
\makeatother

\begin{figure}[t!]
    \centering
    \includegraphics[width=0.5\textwidth]{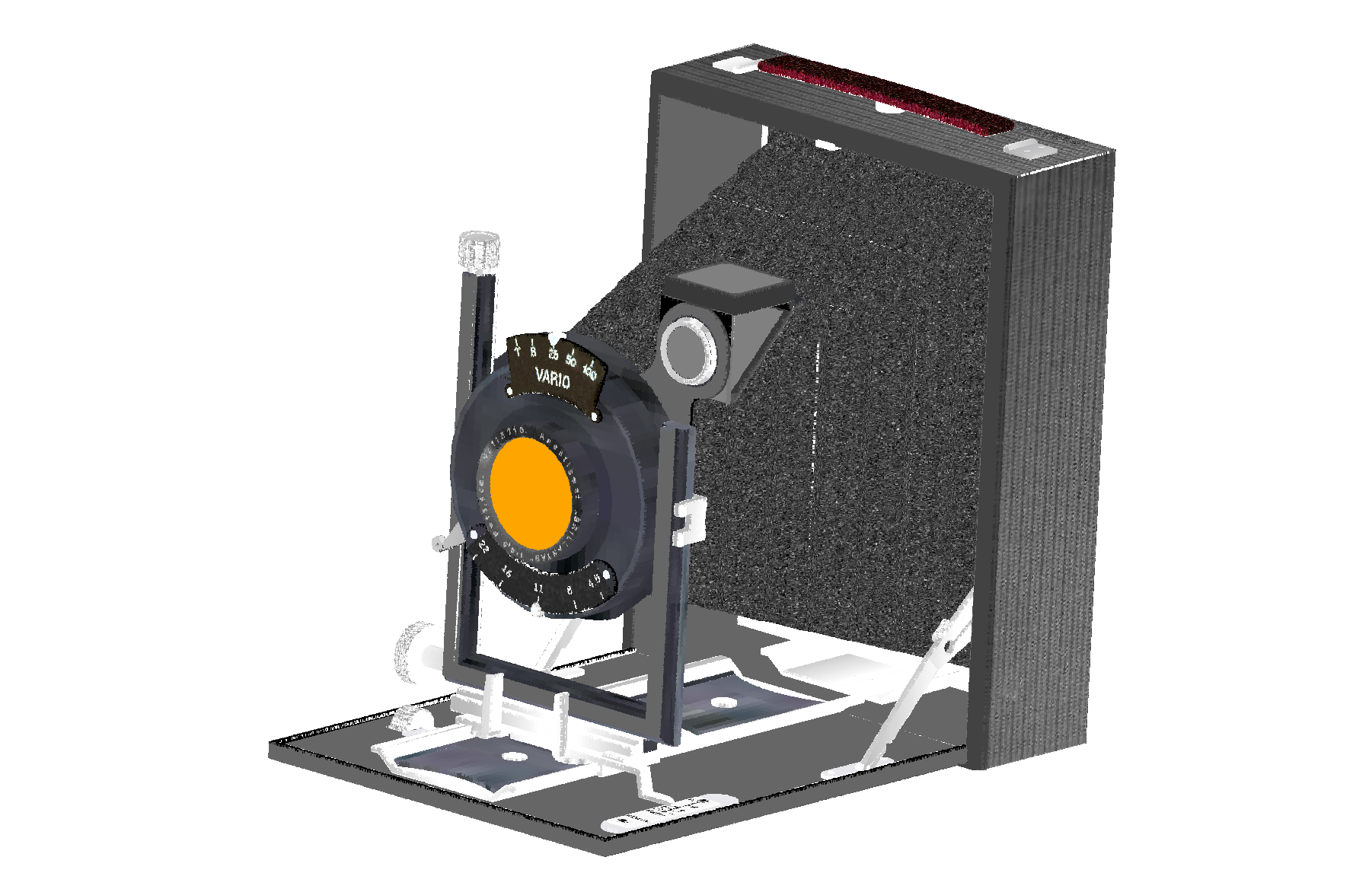}
    \caption{Point cloud generated from \emph{CameraVarioPlatterncamera} by proposed script}
    \label{fig:script_pc}
\end{figure}

\pagebreak

\end{document}